\documentclass[aps,prl,reprint,groupedaddress]{revtex4-1}
\usepackage{graphicx} % Include figure files
\usepackage{dcolumn} % Align table columns on decimal point
\usepackage{bm}% bold math
\usepackage{hyperref}% add hypertext capabilities

\begin{document}

\title{Dynamics of strangeness production in heavy-ion collisions near threshold energies}

\author{Zhao-Qing Feng}
\email{fengzhq@impcas.ac.cn}
\author{Gen-Ming Jin}
\affiliation{Institute of Modern Physics, Chinese Academy of
Sciences, Lanzhou 730000, People's Republic of China}

\date{\today}

\begin{abstract}
Within the framework of the improved isospin dependent quantum
molecular dynamics (ImIQMD) model, the dynamics of strangeness (K$^{0,+}$, $\Lambda$ and $\Sigma^{-,0,+}$) production in
heavy-ion collisions near threshold energies is investigated
systematically, in which the strange particles are considered to be mainly produced by the inelastic collisions of baryon-baryon and pion-baryon.
The collisions in the region of supra-saturation densities of the dense baryonic matter formed in heavy-ion collisions dominate the yields of strangeness production.
The total multiplicities as functions of incident energies and collision centralities are calculated with the Skyrme parameter SLy6.
The excitation function of strangeness production is analyzed and also compared with the KaoS data for the K$^{+}$ production in the reactions $^{12}$C+$^{12}$C and $^{197}$Au+$^{197}$Au.
\begin{description}
\item[PACS number(s)]
25.75.-q, 13.75.Jz, 25.80.Dw
\end{description}
\end{abstract}

\maketitle

Particle production in relativistic heavy-ion collisions has been investigated as a useful tool to extract the information of the nuclear equation of state (EoS) of isospin asymmetric nuclear matter under extreme conditions in terrestrial laboratories, such as high density, high temperature and large isospin asymmetry etc. Besides nucleonic observables such as rapidity distribution and flow of free nucleons or light clusters
(such as deuteron, triton and alpha etc.), also mesons emitted from
the reaction zone can be probes of the hot and dense nuclear matter. In particular, the strangeness productions (K$^{0,+}$, $\Lambda$ and $\Sigma^{-,0,+}$) in heavy-ion collisions in the region of 1A GeV energies have been extensively investigated both experimentally \cite{Fo07,Me07} and theoretically \cite{Ai85,Li97,Fu01,Ha06}. Kaons ($K^{0}$ and $K^{+}$) as a probe of EoS are produced in the high density phase without subsequent reabsorption process. It was noticed that the kaon yields are sensitive to the EoS in theoretical investigations by transport models. The available data already favored a soft EoS at high densities. The $K^{0}/K^{+}$ ratio was also proposed as a sensitive probe to constrain the high-density behavior of the symmetry energy \cite{Fe06,Pr10}. Up to now, the difference of the symmetry energy at supra-saturation densities predicted by transport models is huge.
It is not only in understanding the reaction dynamics, the high-density behavior of the symmetry energy also has an important application in astrophysics,
such as the structure of neutron star, the cooling of protoneutron stars, the nucleosynthesis
during supernova explosion of massive stars etc \cite{St05}.
With the establishment of new-generation radioactive beam
facilities in the world, such as the CSR (IMP in Lanzhou, China),
FAIR (GSI in Darmstadt, Germany), RIKEN (Japan), SPIRAL2 (GANIL in
Caen, France) and FRIB (MSU, USA), the high-density
behavior of the symmetry energy is being studied more detail
experimentally. Precise description of the experimental data is still
necessary by improving transport models or developing some new
approaches, especially for constraining the high-density behavior of the nuclear symmetry energy.

The ImIQMD model has been successfully applied to treat
dynamics in heavy-ion fusion reactions near Coulomb barrier and also to describe the capture of two heavy colliding nuclides
\cite{Fe05,Fe08}. The ground state properties of a single nuclide is described well in the ImIQMD model. Further improvements of the ImIQMD model have been performed in order to investigate the pion dynamics in heavy-ion collisions and to extract the information of isospin asymmetric EoS at supra-saturation densities \cite{Fe09,Fe10,Fg10}. The strangeness productions at near threshold energies are to be included in inelastic baryon-baryon and pion-baryon collisions and investigated in this work.

In the ImIQMD model, the time evolutions of the baryons and mesons in
the system under the self-consistently generated mean-field are
governed by Hamilton's equations of motion, which read as
\begin{eqnarray}
\dot{\mathbf{p}}_{i}=-\frac{\partial H}{\partial\mathbf{r}_{i}},
\quad \dot{\mathbf{r}}_{i}=\frac{\partial
H}{\partial\mathbf{p}_{i}}.
\end{eqnarray}
Here we omit the shell correction part in the Hamiltonian $H$ as
described in Ref. \cite{Fe08}. The Hamiltonian of baryons consists
of the relativistic energy, the effective interaction potential energy and
the momentum dependent part as follows:
\begin{equation}
H_{B}=\sum_{i}\sqrt{\textbf{p}_{i}^{2}+m_{i}^{2}}+U_{int}+U_{mom}.
\end{equation}
Here the $\textbf{p}_{i}$ and $m_{i}$ represent the momentum and the
mass of the baryons. The momentum dependent term is taken as the same form in Ref. \cite{Ai87}, which reduces the effective mass in nuclear medium.

The effective interaction potential is composed of the Coulomb
interaction and the local interaction
\begin{equation}
U_{int}=U_{Coul}+U_{loc}.
\end{equation}
The Coulomb interaction potential is calculated by
\begin{equation}
U_{Coul}=\frac{1}{2}\sum_{i,j,j\neq
i}\frac{e_{i}e_{j}}{r_{ij}}erf(r_{ij}/\sqrt{4L})
\end{equation}
where the $e_{j}$ is the charged number including protons and
charged resonances. The $r_{ij}=|\mathbf{r}_{i}-\mathbf{r}_{j}|$ is
the relative distance of two charged particles.

The local interaction potential energy is derived directly from the Skyrme
energy-density functional and expressed as
\begin{equation}
U_{loc}=\int V_{loc}(\rho(\mathbf{r}))d\mathbf{r}.
\end{equation}
The local potential energy-density functional reads
\begin{eqnarray}
V_{loc}(\rho)=&& \frac{\alpha}{2}\frac{\rho^{2}}{\rho_{0}}+
\frac{\beta}{1+\gamma}\frac{\rho^{1+\gamma}}{\rho_{0}^{\gamma}}+
\frac{g_{sur}}{2\rho_{0}}(\nabla\rho)^{2}  \nonumber \\
&& + \frac{g_{sur}^{iso}}{2\rho_{0}}[\nabla(\rho_{n}-\rho_{p})]^{2}  \nonumber \\
&& + \left(a_{sym}\frac{\rho^{2}}{\rho_{0}}+b_{sym}\frac{\rho^{1+\gamma}}{\rho_{0}^{\gamma}}+
c_{sym}\frac{\rho^{8/3}}{\rho_{0}^{5/3}}\right)\delta^{2}  \nonumber \\
&& + g_{\tau}\rho^{8/3}/\rho_{0}^{5/3},
\end{eqnarray}
where the $\rho_{n}$, $\rho_{p}$ and $\rho=\rho_{n}+\rho_{p}$ are
the neutron, proton and total densities, respectively, and the
$\delta=(\rho_{n}-\rho_{p})/(\rho_{n}+\rho_{p})$ is the isospin
asymmetry. Here, all the terms in the Skyrme energy functional are included in the model besides the spin-orbit coupling. The coefficients $\alpha$, $\beta$, $\gamma$, $g_{sur}$,
$g_{sur}^{iso}$, $g_{\tau}$ are related to the Skyrme parameters
$t_{0}, t_{1}, t_{2}, t_{3}$ and $x_{0}, x_{1}, x_{2}, x_{3}$ as \cite{Fe08},
\begin{eqnarray}
&& \frac{\alpha}{2}=\frac{3}{8}t_{0}\rho_{0}, \quad
\frac{\beta}{1+\gamma}=\frac{t_{3}}{16}\rho_{0}^{\gamma}, \\
&& \frac{g_{sur}}{2}=\frac{1}{64}(9t_{1}-5t_{2}-4x_{2}t_{2})\rho_{0}, \\
&& \frac{g_{sur}^{iso}}{2}=-\frac{1}{64}[3t_{1}(2x_{1}+1)+t_{2}(2x_{2}+1)]\rho_{0}, \\
&& g_{\tau}=\frac{3}{80}\left(\frac{3}{2}\pi^{2}\right)^{2/3}(3t_{1}+5t_{2}+4x_{2}t_{2})\rho_{0}^{5/3}.
\end{eqnarray}
The parameters of the potential part in the bulk symmetry
energy term are also derived directly from Skyrme energy-density
parameters as
\begin{eqnarray}
&& a_{sym}=-\frac{1}{8}(2x_{0}+1)t_{0}\rho_{0}, \quad b_{sym}=-\frac{1}{48}(2x_{3}+1)t_{3}\rho_{0}^{\gamma}, \nonumber \\
&& c_{sym}=-\frac{1}{24}\left(\frac{3}{2}\pi^{2}\right)^{2/3}\rho_{0}^{5/3}[3t_{1}x_{1}-t_{2}(5x_{2}+4)].
\end{eqnarray}
In this work, we use the Skyrme force Sly6 in the ImIQMD calculations, which gives the modulus of incompressibility of symmetric nuclear matter about 230 MeV.

Analogously to baryons, the evolution of mesons (here mainly pions and kaons) is also determined by the Hamiltonian, which is given by
\begin{eqnarray}
H_{M}&& = \sum_{i=1}^{N_{M}}\left( V_{i}^{\textrm{Coul}} + \omega(\textbf{p}_{i},\rho_{i}) \right).
\end{eqnarray}
Here the Coulomb interaction is given by
\begin{equation}
V_{i}^{\textrm{Coul}}=\sum_{j=1}^{N_{B}}\frac{e_{i}e_{j}}{r_{ij}},
\end{equation}
where the $N_{M}$ and $N_{B}$ are the total numbers of mesons and
baryons including charged resonances. Here, we use the energy in vacuum for mesons, namely $\omega(\textbf{p}_{i},\rho_{i})=\sqrt{\textbf{p}_{i}^{2}+m_{M}^{2}}$ with the momentum $\textbf{p}_{i}$ and the mass $m_{M}$ of the mesons.

A hard core scattering in two particle collisions is assumed in the simulation of the collision processes by Monte Carlo procedures, in which the scattering of two particles is determined by a geometrical minimum distance criterion $d\leq\sqrt{0.1\sigma_{tot}/\pi}$ fm weighted by the Pauli blocking of the final states \cite{Ai91,Be88}. Here, the total cross section $\sigma_{tot}$ in mb is the sum of the elastic and all inelastic cross section. The probability reaching a channel in a collision is calculated by its contribution of the channel cross section to the total cross section as $P_{ch}=\sigma_{ch}/\sigma_{tot}$. The choice of the channel is done randomly by the weight of the probability. The primary products in nucleon-nucleon (NN) collisions at the 1A GeV energies are the resonances $\triangle$(1232), N*(1440) and the pions. The reaction channels are given as follows:
\begin{eqnarray}
&& NN \leftrightarrow N\triangle, \quad  NN \leftrightarrow NN^{\ast}, \quad  NN
\leftrightarrow \triangle\triangle,  \nonumber \\
&& \Delta \leftrightarrow N\pi,  N^{\ast} \leftrightarrow N\pi,  NN \rightarrow NN\pi (s-state).
\end{eqnarray}
The cross sections of each channel to produce resonances are
parameterized by fitting the data calculated with the one-boson
exchange model \cite{Hu94}. In the 1 A GeV region, there are mostly
$\Delta$ resonances which disintegrate into a $\pi$ and a nucleon,
however, the $N^{\ast}$ yet gives considerable contribution to the
high energetic pion yield. The energy and momentum dependent decay
width is used in the calculation \cite{Fe09}. The strangeness is created by inelastic hadron-hadron collisions. We include the
channels as follows:
\begin{eqnarray}
&& BB \rightarrow BYK,  BB \rightarrow BBK\overline{K},  B\pi \rightarrow YK,  B\pi \rightarrow NK\overline{K}, \nonumber \\
&& Y\pi \rightarrow N\overline{K}, \quad  N\overline{K} \rightarrow Y\pi, \quad YN \rightarrow \overline{K}NN.
\end{eqnarray}
Here the B strands for (N, $\triangle$) and Y($\Lambda$, $\Sigma$), K(K$^{0}$, K$^{+}$) and $\overline{K}$($\overline{K^{0}}$, K$^{-}$). The parameterized cross sections of each isospin channels $BB \rightarrow BYK$ \cite{Ts99} are used in the calculation. We take the parametrizations of the channels $B\pi \rightarrow YK$ \cite{Ts94} besides the $N\pi \rightarrow \Lambda K$ reaction \cite{Cu84}. The results are close to the experimental data at near threshold energies. The cross section of antikaon production in inelastic hadron-hadron collisions is taken as the same form of the parametrization used in hadron string dynamics (HSD) calculations \cite{Ca97}. Furthermore, the elastic channels are considered through the channels $KB \rightarrow KB$ and $\overline{K}B \rightarrow \overline{K}B$ and we use the parametrizations in Ref. \cite{Cu90}.

\begin{figure*}
\includegraphics{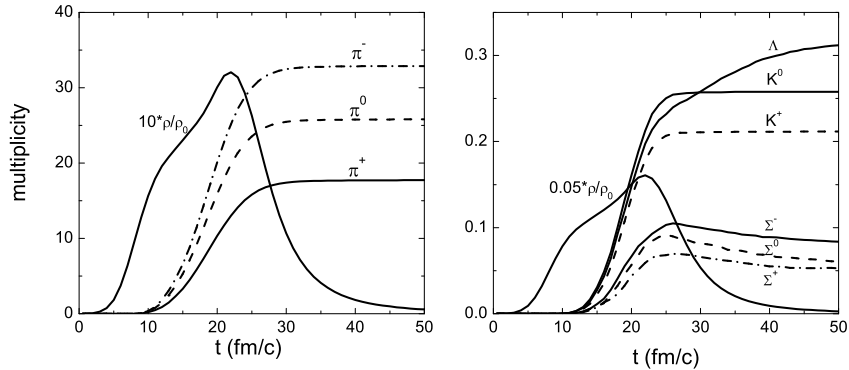}
\caption{\label{fig:wide} Time evolution of the central density, pion and strangeness produced in the reaction $^{197}$Au+$^{197}$Au for head-on collisions at incident energy 1.5A GeV.}
\end{figure*}

We calculated the production of pions, kaons and hyperons and compared with the evolutions of the central density in central $^{197}$Au+$^{197}$Au collisions at incident energy 1.5A GeV as shown in Fig. 1. One can see that the pions and strange particles are mainly produced at supra-saturation densities ($\rho>\rho_{0}$). The observables can be probes to extract the high-density information of EoS. The pion yields saturate at time of the order of 30 fm/c, and the kaons are produced at early stage of the reaction around 25 fm/c. The results are similar to the reports in Ref. \cite{Fe06} but more pronounced in the RBUU calculations. The $B\pi \rightarrow YK$ contributes to be about 1/3 of the total kaon production, which retards the profile of the kaon yields in the calculations. The saturation of the hyperons is more slowly than the cases of pions and kaons owing to the exchange reactions $Y\pi \rightarrow N\overline{K}$ and $N\overline{K} \rightarrow Y\pi$. The isospin difference is pronounced in the temporal evolutions, which enables the observables as probes of the high-density behavior of the symmetry energy, such as the ratios $\pi^{-}/\pi^{+}$, K$^{0}$/K$^{+}$ and $\Sigma^{-}/\Sigma^{+}$.

\begin{figure*}
\includegraphics{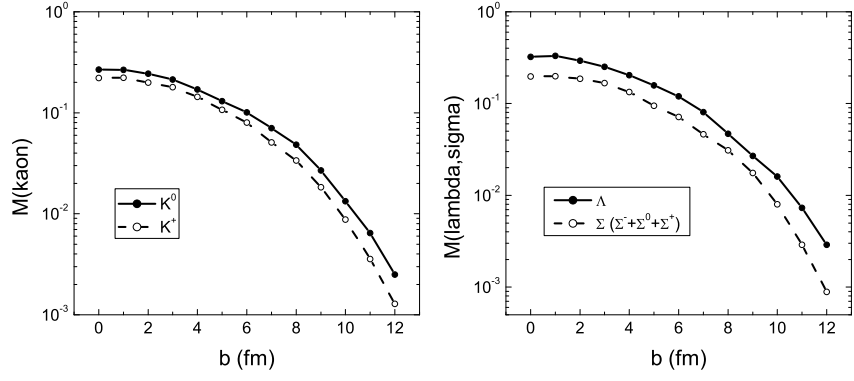}
\caption{\label{fig:wide} Total multiplicities of the final K$^{0,+}$, $\Lambda$ and the sum of $\Sigma$ as a function of impact parameter in the reaction $^{197}$Au+$^{197}$Au at 1.5A GeV.}
\end{figure*}

\begin{figure*}
\includegraphics{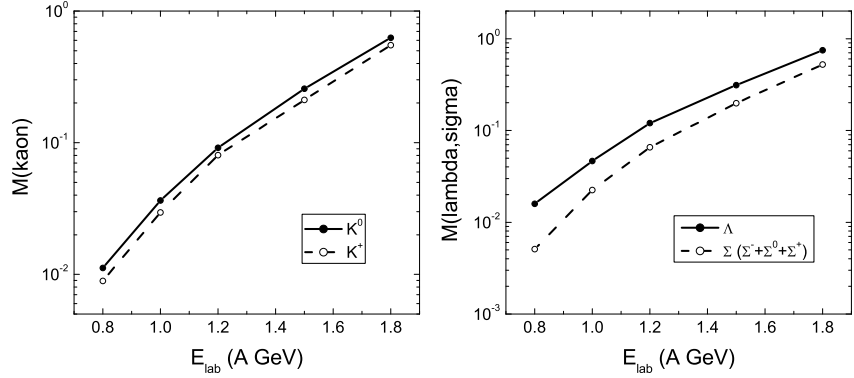}
\caption{\label{fig:wide} The same as in Fig. 2, but the calculated excitation functions of strangeness production for head-on collisions.}
\end{figure*}

The strange particles in heavy-ion collisions are mainly created at supra-saturation densities. To extract the basic information of the compressed nuclear matter at high densities formed in relativistic heavy-ion collisions, the centrality and incident energy dependence of the strangeness production was calculated as shown in Fig. 2 and in Fig. 3 for the $^{197}$Au+$^{197}$Au reaction. A larger domain of the high-density phase diagram is formed in central collisions and with increasing incident energy, which enhances the probabilities of NN inelastic collisions and hence the strangeness production. The neutron-neutron collisions contribute the productions of K$^{0}$ and $\Sigma^{-}$. Therefore, the neutron-rich systems augment the values of the K$^{0}$/K$^{+}$ and $\Sigma^{-}/\Sigma^{+}$ ratios.

\begin{figure*}
\includegraphics{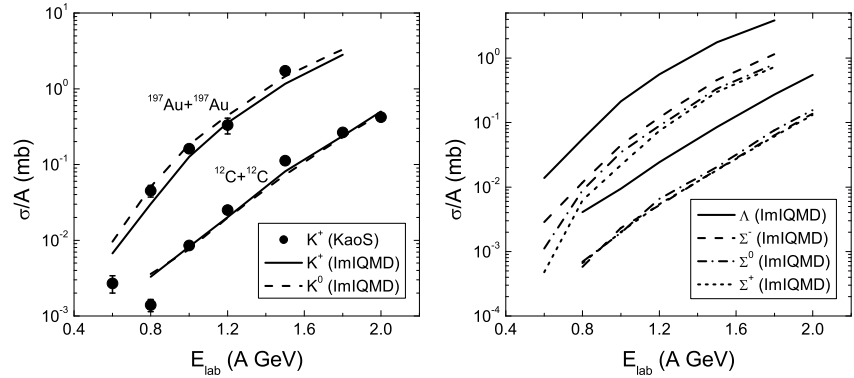}
\caption{\label{fig:wide} Comparison of the calculated strangeness production cross section per mass number ($\sigma$/A) and the KaoS data for the K$^{+}$ production for the system $^{197}$Au+$^{197}$Au and the reaction $^{12}$C+$^{12}$C.}
\end{figure*}

The reliability of the calculations on the strangeness production can be checked from the available experimental data. Shown in Fig. 4 is the calculated excitation functions of strange particles for the heavy $^{197}$Au+$^{197}$ and the light $^{12}$C+$^{12}$C reactions and compared with the KaoS data for the K$^{+}$ production \cite{Fo07}. The experimental data can be well reproduced by calculations besides at very low threshold energies owing to the limited statistics. The larger cross sections of the strangeness production for heavy systems result from the larger region of the high-density phase diagram. The difference of production of the same isospin particles such as K$^{0,+}$ or $\Sigma^{-,0,+}$ is deviated from the isospin effects of collision systems. Influence of the symmetry energy and the in-medium potentials on the strangeness production is being investigated.

In summary, the production of strangeness in heavy-ion collisions in the region of 1A GeV energies is investigated within the framework of the ImIQMD model. The strange particles are produced at the supra-saturation densities and emitted at the early stage of the phase diagram comparing with the pion production. Strangeness exchange process retards the saturation of $\Lambda$ and $\Sigma$ production. Higher incident energy and central collisions enhance the domain of the high-density phase diagram, hence lead to the increase of the strangeness multiplicity. A larger high-density region of compressed nuclear matter is formed in heavy collision systems comparing with the light systems, which increases the strangeness production. The experimental data of the K$^{+}$ production cross sections at near threshold energies can be reproduced rather well.

\bigskip
\begin{acknowledgments}
This work was supported by the National Natural Science Foundation
of China under Grant 10805061; the Special Foundation of the
President Fund; the West Doctoral Project of Chinese Academy of
Sciences; and the Major State Basic Research Development Program
under Grant 2007CB815000.
\end{acknowledgments}

\end{document}